\documentstyle[eqsecnum,manuscript,aps,psfig]{revtex}
\def\be{\begin{equation}}
\def\ee{\end{equation}}
\def\bea{\begin{eqnarray}}
\def\eea{\end{eqnarray}}

\begin{document}
\title{Energy-momentum and angular momentum of G\"odel Universes}
\author{Mariusz P. D\c abrowski\footnote{E-mail:mpdabfz@uoo.univ.szczecin.pl} and
Janusz Garecki\footnote{E-mail:garecki@wmf.univ.szczecin.pl}}
\address{Institute of Physics, University of Szczecin, 70-451 Szczecin, Poland.}
\date{\today}
\maketitle

\begin{abstract}
We discuss the Einstein energy-momentum complex and the Bergmann-Thomson
angular momentum complex in general relativity and calculate them
for space-time homogeneous
G\"odel universes. The calculations are performed for a dust
acausal G\"odel model and for a scalar-field causal
G\"odel model. It is shown that the Einstein pseudotensor is traceless,
not symmetric,
the gravitational energy "density" is negative and that the gravitational Poynting
vector vanishes. Significantly, the total (gravitational and matter) energy
"density" for the acausal model is zero while for the causal model it is negative.
The Bergmann-Thomson angular momentum complex does not vanish for
both G\"odel models.

\end{abstract}

%\preprint{hep-th/0309064}

PACS number(s): 98.80.Hw, 98.80.-k,98.80.Jk,04.20.Cv

\newpage

\section{Introduction}
\label{sect1}

\setcounter{equation}{0}

The problem of the energy-momentum of gravitational field has a
very long tradition in general relativity. The point is that
the gravitational field can be made locally vanish and so one is
always able to find the frame in which the energy-momentum of gravitational
field is zero
while in the other frame it is not. In other words, the physical
objects which can describe this situation cannot be tensors, i.e.,
the objects which vanish in all the frames provided they vanish in
at least one of them. The proposed quantities which actually
fulfill the conservation law of matter appended with gravitational
field are called energy-momentum {\it complexes} while their
gravitational parts are called gravitational field {\it pseudotensors}.
An energy-momentum complex is then the sum of the obvious
energy-momentum tensor of matter and an appropriate pseudotensor.
Unfortunately, the choice of the gravitational field pseudotensor
is not unique and because of that quite a few definitions of these
pseudotensors
have been proposed. Historically, one of the earliest definitions
was given by Einstein followed by Landau-Lifshitz \cite{landau},
M\o ller \cite{Mol72}, Papapetrou \cite{papapetrou}, Bergmann-Thomson \cite{BT},
Weinberg \cite{Weinberg} and Bak-Cangemi-Jackiw \cite{Jackiw}, for
example. Among them only those of Landau-Lifshitz, Weinberg
and Bak-Cangemi-Jackiw are symmetric, but only in holonomic frames. 
In particular, the Einstein
pseudotensor is not symmetric. The problem of the energy-momentum
of the gravitational field can also be extended to the standard
field theory problem of the angular momentum. The appropriate
expressions have been proposed of which the Bergmann-Thomson angular
momentum complex \cite{BT} being the most widely used.

Because of the freedom of a choice of pseudotensors and the
fact that they usually give {\it different} results for the same type of
spacetime some authors \cite{supertensors,gar77-96,gar99-02} have proposed
an alternative approach to
the problem in which they defined the quantities which describe the
generalized energy-momentum content of the gravitational field and which are {\it
tensors}. These quantities are called gravitational
superenergy tensors and gravitational supermomentum tensors.

It seems interesting to make a comparative analysis of the results which
can be obtained in the energy pseudotensor approach with that of the superenergy
tensor approach for variuos models of spacetime. The question arises
whether the appropriate physical conclusions obtained on the level
of the energy-momentum are preserved on the level of the supermomentum and
vice versa.

In this context the canonical superenergy tensor and
the canonical angular supermomentum tensor for space-time homogeneous universes
of G\"odel type have been calculated and discussed by these authors
recently \cite{paper1,godel,metric}.
The task of this paper is to calculate the appropriate
pseudotensors (complexes) and make the comparison of the physical
results.

It is not random that we have chosen G\"odel universes as the
example models to compare the results. Firstly, G\"odel universes rotate and so
they should have non-zero angular momentum. Secondly, they possess closed
timelike curves (CTCs) which is a big peculiarity and may have interesting
consequences onto the results. In particular, the CTCs
should be avoided according to the Hawking's chronology protection
conjecture \cite{hawking} and this somehow may be related to the energy-momentum
and the angular momentum in the same way as it was the case for the superenergy
and the supermomentum in Ref.\cite{paper1}.

On the other hand, following Hawking's chronology protection conjecture
it has been shown that it is possible to avoid CTCs in many gravitational theories.
This is the case in minimally coupled to gravity
scalar field theories \cite{tiomno}, in quadratic gravity theories \cite{accio},
in five-dimensional gravity theories
\cite{teixeira}, or in string/M-theory inspired gravitational theories
\cite{bardab98,new,barrow}. In Ref. \cite{barrow}, for example, it
has been shown that CTCs can be avoided for brane models with the negative
total effective energy density. One should also emphasize that
G\"odel universes attracted attention of many authors recently,
just in the context of conventional gravity theory
\cite{BonOzs,Gad}.

In this paper we study field theoretical quantities such as the energy-momentum
and the angular momentum for G\"odel universes. In order to fulfill the
task we apply the Einstein energy-momentum complex of gravitation
and matter and the Bergmann-Thomson angular momentum complex.
These quantities seem to be the best of all which have been
proposed so far, including the so-called ``quasi-local
quantities''. We perform our analysis in orthonormal frames
(anholonomic frames) which requires adopting the original
expressions to these frames. The obvious way to express
Einstein complex and Bergmann-Thomson complex in anholonomic
frames is to use the formalism of the tensor-valued (or
pseudotensor-valued) exterior differential forms which we do in
Section II. In Section III we apply the obtained quantities for
G\"odel spacetimes. In Section IV we conclude and make some comparison of
energetic quantities with superenergetic quantities for G\"odel
spacetimes which have been obtained earlier \cite{paper1}.

\section{Energy-momentum and angular momentum in general relativity}
\label{sect2}

\setcounter{equation}{0}

As it was already mentioned in the Introduction the gravitational field
{\it does not possess} the proper definition of an energy-momentum tensor
and an angular momentum tensor and one usually defines some
energy-momentum pseudotensors. The thorough investigations of the energy-momentum
problem in general relativity suggest \cite{Mol72,Atretal}
that the most satisfactory of all the possible gravitational energy pseudotensors
already listed in the Introduction is the canonical gravitational energy-momentum
pseudotensor of Einstein $_E t_i^{~k}$ (see e.g. \cite{landau}). In consequence,
the best of all the proposed
gravitational angular momentum pseudotensors is considered to be the
Bergmann-Thomson pseudotensor \cite{BT} since it is constructed of
the Einstein pseudotensor. We follow this point of view and
will discuss a particular application of these pseudotensors to G\"odel universes.

Independently, in general relativity one can also introduce the
canonical gravitational superenergy tensor and the canonical
gravitational angular supermomentum tensor. This was done in a
series of one of the authors' papers \cite{gar77-96,gar99-02}.
It appeared that the idea of
the superenergy and the angular supermomentum tensors was
universal: {\it to any physical field which possesses an
energy-momentum tensor or a pseudotensor which is constructed out of
the Levi-Civita connection one can always attribute a
corresponding superenergy tensor and a corresponding angular supermomentum tensor}.

The canonical superenergy and angular supermomentum tensors prove
very useful for the local analysis of the gravitational and
matter fields. They also admit suitable global integral
superenergetic quantities for gravity and matter \cite{gar99-02}.

In this paper we confine ourselves to the analysis of the
energetic quantities for G\"odel spacetimes. In fact, we calculate
the energy-momentum ``densities'' and the angular momentum
``densities'' for these spacetimes. In order to calculate these
quantities we use the expressions for Einstein energy-momentum
pseudotensor and Bergmann-Thomson angular momentum complex in
an anholonomic form. The appropriate formulas which are valid in an
arbitrary frame $(\theta^i)$ can be obtained by the application of the
tensor-valued differential forms \cite{Thir78,Atr84}.

In the language of the differential forms the Einstein equations
read as
\be
\label{diffEE}
\frac{1}{2} \Omega^{j}_{~k} \wedge \eta_{ij}^{~~k} = - \chi T_i ,
\ee
where
\be
\Omega^{k}_{~l} = \frac{1}{2} R^{k}_{~lmn} e^m \wedge e^n
\ee
is the curvature 2-form of the Riemannian (or Levi-Civita)
connection 1-form $\omega^{i}_{k} = \Gamma^{i}_{kl} \theta^l$, $\chi = 8\pi$ $(G = c =1)$,
and
\be
\eta_{ij}^{~~k} = g^{kl} \eta_{ijl} = g^{kl} e^r \eta_{ijlr} = g^{kl} e^r \sqrt{\mid g \mid}
\epsilon_{ijlr}
\ee
is a pseudotensorial 1-form with $\epsilon_{ijlr}$ being the totally antisymmetric
Levi-Civita pseudotensor. In the following we will use an anholonomic Lorentzian
frame $(e^i)$ defined by
\be
\label{coreper}
g = \eta_{ik} e^i \otimes e^k ,
\ee
where $g$ is an arbitrary spacetime metric and $\eta_{ik}$ is
Minkowski metric. In (\ref{diffEE}) $T_i := T_{i}^{~k} \eta_k$ is the energy-momentum
3-form of matter with $T_{i}^{k}$ being the symmetric energy-momentum tensor
of matter, and
\be
\eta_i = \frac{1}{3} e^i \wedge \eta_{ij} = \frac{1}{6} e^j \wedge
e^k \wedge \eta_{ijk}
\ee
is a pseudotensorial 3-form.

Decomposing (\ref{diffEE}) in the basis of the 3-forms $\eta_i$,
one can easily get the Einstein equations in an ordinary tensorial
form
\be
G_{ik} = \chi T_{ik} ,
\ee
where
\be
G_{ik} := R_{ik} - \frac{1}{2} g_{ik} R
\ee
are the components of the Einstein tensor. It is known that the
Einstein equations (\ref{diffEE}) can also be transformed to the
superpotential form
\be
\label{EEsuperp}
d\left( \frac{1}{2\chi} \eta_{ij}^{~~k} \wedge \omega^{j}_{~k}
\right) = T_i + \frac{1}{2\chi} \left( \eta_{pj}^{~~k} \wedge
\omega^{j}_{~k} \wedge \omega^{p}_{~i} +
\eta_{ij}^{~~p} \wedge \omega^{k}_{~p} \wedge \omega^{j}_{~k} \right).
\ee

The equations (\ref{EEsuperp}) are independent of coordinates (or
frames) and define the canonical 3-form of the gravitational
energy-momentum
\be
_E t_i := \frac{1}{2\chi} \left( \eta_{pj}^{~~k} \wedge
\omega^{j}_{~k} \wedge \omega^{p}_{~i} +
\eta_{ij}^{~~p} \wedge \omega^{k}_{~p} \wedge \omega^{j}_{~k} \right) ,
\ee
and the 2-form
\be
_F U_i := \frac{1}{2\chi} \eta_{ij}^{~~k} \wedge \omega^{j}_{~k}
\ee
gives the so-called Freud superpotentials.

The sum
\be
\label{Ecomplex}
_E t_i + T_i := _E K_i
\ee
composes the 3-form $_E K_i$ which we call the canonical
Einstein energy-momentum complex of gravitation and matter.
From (\ref{EEsuperp}) and (\ref{Ecomplex}) we have
\be
\label{KdU}
_E K_i = d_F U_i .
\ee
A troublesome fact is that the 3-form $_E t_i$ and, in
consequence, the 3-forms $d_FU_i$ and $_E K_i$ are non-tensorial.
This means gravitational energy-momentum is not localizable. In
fact, only the global energy-momentum can be properly defined in
the asymptotically flat spacetimes (at null and spatial infinity).
From (\ref{KdU}) one immediately gets the local, differential
energy-momentum conservation laws for gravity and matter (also
called weak conservation laws -- they hold in any reference frame) in the form
\be
\label{dEK}
d_E K_i = 0 .
\ee
The integration of (\ref{dEK}) over a compact 4-dimensional domain
$\Omega$ leads to Synge's integral conservation laws \cite{synge}
\be
\label{intdEK}
\int_{\partial V} \left( _E t_i + T_i \right) = 0 ,
\ee
where $\partial V$ denotes a 3-dim outward oriented boundary of the
4-dim domain $V$.

It is interesting to note that the integrals on the left-hand side
of (\ref{intdEK}) have no geometrical meaning, but they are zero
in any reference frame, i.e., they behave like scalars. Moreover,
for a closed system \cite{Mol72}, after the
appropriate choice of the domain $V$ one obtains from (\ref{intdEK}) the ordinary
conservation laws for energy-momentum of matter and gravitation.

In the basis of the 3-forms $\eta_i$ one can decompose the
canonical 3-form of the gravitational energy-momentum as follows
\be
_E t_i = _E t_i^{~q}\eta_{q},
\ee
and its components form the energy-momentum pseudotensor of
Einstein.
In a Lorentzian frame $(e^i)$ we have
\be
\label{Epseudo}
_E t_{i}^{~q} = \frac{1}{2\chi} \left( g^{kl} \eta^{qtrs}
\eta_{pjlr} \gamma^{j}_{kt} \gamma^{p}_{is} +
g^{pl} \eta^{qtrs} \eta_{ijlr} \gamma^{k}_{pt} \gamma^{j}_{ks}
\right),
\ee
where $\gamma$'s denote Ricci rotation coefficients,i.e., Levi-Civita connection
in this frame. Let us also mention that in a Lorentzian frame $(e^i)$ one has
$g = -1, \eta^{0123} = 1, \eta_{0123} = -1, g^{ik} = \eta^{ik},
g_{ik} = \eta_{ik}$.

In section \ref{sect3} we will use the formula (\ref{Epseudo}) to calculate the
energy-momentum ``densities'' for G\"odel spacetimes in an
appropriate Lorentzian frame.

Now we turn into the problem of angular momentum in general
relativity which is more complicated than the problem of
energy-momentum (see e.g. \cite{winicour}). The main new obstacle is
that the coordinates $(x^i)$ do not form the components of any
global radius vector $\vec{r}$ so even an ordinary field
theoretical matter angular momentum
\be
\label{FTangmom}
_m M^{ika} = \sqrt{\mid g \mid} \left( x^i T^{ka} - x^k T^{ia} \right)
\ee
does not form a tensor density. In general relativity one can
define the radius vector only locally. For example, the normal
coordinates $(y^i)$ form the components of the local radius vector
$\vec{r}$ with respect to their origin.

In the following we will define the components $(r^i)$ of the
local radius vector $\vec{r}$ with respect to the Lorentzian
frame $(e^i)$ by
\be
Dr^i = e^i,
\ee
where $D$ is the exterior covariant derivative.

In the normal coordinates at the point P, NC(P), this gives the
equality between the normal coordinates and the local radius
vector
\be
r^i = y^i  .
\ee

Apart from this first obstacle there is another. In
general, it is difficult to define invariantly the angular momentum in an
asymptotically flat spacetimes and also the resulting global
angular momentum integrals in radiative spacetimes do not converge (see e.g. \cite{winicour}).
However, these problems can be avoided in the case of closed
systems provided one applies e.g., the definition of the angular momentum
given by Bergmann and Thomson. This is what we now call the
Bergmann-Thomson angular momentum complex. Because of the fact
that it is closely related to the Einstein energy-momentum complex we
call it canonical, too. Using the Bergmann-Thomson
angular-momentum complex, one can also reflect the temporal changes
of the global angular momentum in asymptotically flat spacetimes
\cite{cresswell}.

Bearing in mind all the arguments for the Bergmann-Thomson angular
momentum complex we will apply this complex to calculate
angular momentum densities for G\"odel spacetimes in a Lorentzian
frame $(e^i)$. In order to get a suitable formula in a Lorentzian
frame we start with equations (\ref{Ecomplex}) and (\ref{KdU}) with raised index $i$ to get
\be
\label{FTBT1}
r^i \left( _E t^k + T^k \right) - r^k \left( _E t^i + T^i \right)
= r^i d_F U^k - r^k d_F U^i ,
\ee
or
\be
\label{FTBT2}
r^i \left( _E t^k + T^k \right) - r^k \left( _E t^i + T^i \right)
+ dr^i \wedge _F U^k - dr^k \wedge _F U^i = d \left( r^i _F U^k - r^k _F U^i
\right) .
\ee

The equations (\ref{FTBT2}) hold in any reference frame (both
holonomic and anholonomic) and give the local, differential
conservation laws for the angular momentum of gravitation and
matter
\be
d \left[ r^i \left( _E t^k + T^k \right) - r^k \left( _E t^i + T^i \right)
+ dr^i \wedge _F U^k - dr^k \wedge _F U^i \right] = 0 ,
\ee
and the integral Synge's conservation laws \cite{synge}
\be
\int_{\partial V} \left[ r^i \left( _E t^k + T^k \right) - r^k \left( _E t^i + T^i \right)
+ dr^i \wedge _F U^k - dr^k \wedge _F U^i \right] = 0 .
\ee
The 3-form (\ref{FTBT2}) gives the ``densities'' of the total
canonical angular momentum for gravitation and matter.
Decomposing it in the basis $\eta_i$ one can obtain the antisymmetric
in the first two indices components
$_{BT} M^{ika} = - _{BT} M^{kia}$ of the canonical Bergmann-Thomson
angular momentum complex of gravitation and matter in Lorentzian frames as follows
\be
d \left( r^i d_F U^k - r^k d_F U^i \right) := _{BT} M^{ikl} \eta_l ,
\ee
where
\bea
\label{BTcomplex}
_{BT} M^{ika} = \frac{1}{2\chi} \left[ \eta_{lmrn} g^{tr}
\gamma^{m}_{tp} \left( \eta^{aipn} g^{kl} - \eta^{akpn} g^{il}
\right) + \eta^{atbs} \eta_{lmrs} \gamma^{m}_{nb} g^{nr} r^p
\left( g^{il} \gamma^{k}_{pt} - g^{kl} \gamma^{i}_{pt} \right)
\right. \nonumber \\
+ \left. \left( r^i g^{kl} - r^k g^{il} \right) \eta^{aspn} \left(
\eta_{tjmn} g^{rm} \gamma^{t}_{ls} \gamma^{j}_{rp} - \eta_{ljmn}
g^{rm} \gamma^{t}_{rs} \gamma^{j}_{tp} - \frac{1}{2} \eta_{ljms}
g^{tm} R^{j}_{~tnp} \right) \right] ,
\eea
and as $g^{kl}$ one should take Minkowski metric $\eta^{kl}$.
In order to get gravitational part of this complex only, one should
subtract the material part (\ref{FTangmom}).
In Section \ref{sect3} we will use the
formula (\ref{BTcomplex}) in order to calculate the canonical
angular momentum densities for G\"odel spacetimes.

\section{Energy-momentum and angular momentum complexes of G\"odel universes}
\label{sect3}
\setcounter{equation}{0}

Following Ref. \cite{paper1} we will perform the calculations of
the energy-momentum complex and the angular momentum complex for generalized
G\"odel spacetimes in a Lorentzian frame $(e^i)$ defined by
\begin{eqnarray}
\label{eeee}
e^0 & = & dt' + H(x) dy \nonumber \\
e^1 & = & dx \nonumber \\
e^2 & = & D(x) dy \nonumber \\
e^3 & = & dz ,
\end{eqnarray}
where
\be
H(x) = e^{mx}, \hspace{0.5cm} D(x) = \frac{e^{mx}}{\sqrt{2}} ,
\ee
and $m=$ const. The appropriate line element in the coordinates $(t',x,y,z)$
reads as
\begin{equation}
\label{godelmet}
ds^2= - \left[ dt' + H(x)dy \right]^2 - D(x)^2 dy^2 + dx^2+dz^2 .
\end{equation}
The only non-vanishing Ricci rotation coefficients \cite{paper1} in the
Lorentzian frame (\ref{eeee}) are
\bea
\label{gammas}
\gamma^{0}_{12} = \gamma^{1}_{20} = \gamma^{1}_{02} =
\frac{m}{\sqrt{2}}, \nonumber \\
\gamma^{0}_{21} = \gamma^{2}_{10} = \gamma^{2}_{01} =
- \frac{m}{\sqrt{2}}, \nonumber \\
\gamma^{1}_{22} = - \gamma^{2}_{12} = - m .
\eea
According to (\ref{RT}) and (\ref{G}) one has to put $m = \sqrt{2} \Omega$
for an acausal G\"odel model, and $m = 2\Omega$ for a causal model \cite{tiomno,accio}.

In order to learn about causality one has to make a change of
coordinates from $(t',x,y,z)$ into $(t,r,\psi,z)$ as follows
\bea
x & = & \frac{1}{m} \ln{\left[\cosh{(mr)} + \cos{\psi}\right]}\\
y & = & - \frac{\sqrt{2}}{m} \frac{\sin{\psi}\sinh{(mr)}}{\cosh{(mr)} +
\cos{\psi}} \\
t' & = & t + \frac{\sqrt{2}}{m} \left[2 {\rm arctg}{\left(e^{-mr}
\tan{\frac{\psi}{2}}\right)} - \psi \right] \\
z & = & z
\eea
which brings the metric (\ref{godelmet}) into the form
\be
\label{godelmetc}
ds^2=-dt^2-2H(r)dtd\psi +G(r)d\psi ^2+dr^2+dz^2,
\ee
where
\bea
\label{G}
G(r) = D^2(r) - H^2(r) \equiv \left[ \frac{1}{m} \sinh{(mr)} \right]^2
- \left[ \frac{4\Omega}{m^2} \sinh^2{\left(\frac{mr}{2}\right)} \right]^2 \nonumber \\
= \frac 4{m^2}\sinh ^2{\left( \frac{mr}2\right) }\left[ 1+\left( 1-
\frac{4\Omega ^2}{m^2}\right) \sinh ^2{\left( \frac{mr}2\right) }\right] ,
\eea
with $m$ and $\Omega $ constants. In fact, $m$ is a parameter
which may distinguish between causal and acausal G\"odel
spacetimes. For a perfect-fluid source it is restricted by
\cite{tiomno}
\be
0 \leq m^2 \leq 2 \Omega^2 ,
\ee
while for a scalar field as the source of gravity it has
the values \cite{tiomno}
\be
2 \Omega^2 \leq m^2 \leq 4 \Omega^2 .
\ee
Taking
\be
\label{goed}
m^2 = 2 \Omega^2 ,
\ee
one gets the original G\"odel
spacetime \cite{godel} in which we have an acausal region
and $G(r)$ in Eq. (\ref{G}) becomes negative. This region
appears for a radial coordinate
\be
r > r_c, \hspace{0.7cm} {\rm where} \hspace{0.7cm} \sinh^2{\left( \frac{mr_c}{2}
\right)} = 1 .
\ee
However, in the case of the scalar field source one can take
\be
\label{RT}
4\Omega^2 = m^2,
\ee
which gives
\begin{equation}
\label{causcond}
G(r)=D^2(r)-H^2(r)>0
\end{equation}
and the term in front of $d\psi^2$ in the metric (\ref{godelmetc})
remains positive. The conditions (\ref{RT}) and (\ref{causcond})  remove CTCs to a point
which is formally at $r_c = \infty$. We call the model given by the condition
(\ref{RT}) the causal G\"odel spacetime. In fact, there is a larger class of such causal
G\"odel models \cite{carrion,Reb98} for which there are
no CTCs for any value of the radial coordinate $r > 0$.

The only nonvanishing components of the Riemann tensor in a Lorentzian frame (\ref{eeee})
permitted by the space-time homogeneity of the G\"odel universe are \cite{tiomno,accio}
\begin{equation}
\label{Riemmans}
R_{0101}=R_{0202}=\frac 14\left( \frac{H^{\prime }}D\right) ^2=\Omega ^2,
\hspace{0.5cm}R_{1212}=\frac 34\left( \frac{H^{\prime }}D\right)^2 -\frac{%
D^{\prime \prime }}D=3\Omega^2 - m^2,
\end{equation}
where $m = \sqrt{2} \Omega$ for the acausal model, and
$m = 2 \Omega$ for the causal model, $(\ldots)^{'} = \partial/\partial
x$.

Using (\ref{gammas}) one can easily calculate
the Einstein energy-momentum pseudotensor (\ref{Epseudo}). Its
non-vanishing components are
\be
\label{E}
_E t_0^{~0} = \frac{m^2}{16\pi}, \hspace{0.3cm}
_E t_1^{~1} = - \frac{m^2}{16\pi}, \hspace{0.3cm}
_E t_2^{~2} = -\frac{m^2}{16\pi}, \hspace{0.3cm}
_E t_3^{~3} = \frac{m^2}{16\pi}, \hspace{0.3cm}
_E t_2^{~0} = -\frac{m^2\sqrt{2}}{16\pi}, \hspace{0.3cm}
\ee
which according to (\ref{goed}) and (\ref{RT}) give
\be
\label{Eacaus}
_E t_0^{~0} = \frac{\Omega^2}{8\pi}, \hspace{0.3cm}
_E t_1^{~1} = - \frac{\Omega^2}{8\pi}, \hspace{0.3cm}
_E t_2^{~2} = -\frac{\Omega^2}{8\pi}, \hspace{0.3cm}
_E t_3^{~3} = \frac{\Omega^2}{8\pi}, \hspace{0.3cm}
_E t_2^{~0} = -\frac{\sqrt{2}\Omega^2}{8\pi}, \hspace{0.3cm}
\ee
for an original acausal G\"odel spacetime \cite{godel}, and
\be
\label{Ecaus}
_E t_0^{~0} = \frac{\Omega^2}{4\pi}, \hspace{0.3cm}
_E t_1^{~1} = - \frac{\Omega^2}{4\pi}, \hspace{0.3cm}
_E t_2^{~2} = -\frac{\Omega^2}{4\pi}, \hspace{0.3cm}
_E t_3^{~3} = \frac{\Omega^2}{4\pi}, \hspace{0.3cm}
_E t_2^{~0} = -\frac{\sqrt{2}\Omega^2}{4\pi}, \hspace{0.3cm}
\ee
for a causal G\"odel spacetime \cite{tiomno}. In Ref. \cite{Gad}
the calculation of the Landau-Lifshitz and M\o ller pseudotensors
were performed in holonomic coordinates for the acausal model and they give different results from
ours. However, in the orthonormal frames the Landau-Lifshitz and Eintein pseudotensors 
coincide and the results should be the same (see e.g. \cite{Thir78}).

From (\ref{Eacaus}) and (\ref{Ecaus}) one can conclude that in
both cases the Einstein pseudotensor is traceless, but (as expected)
not symmetric, and that the gravitational energy
``density''
\be
\epsilon_g := _E t_i^{~k} v^i v_k
\ee
is in both cases negative ($v^i = \delta_0^i, v_k = g_{k0}$ for G\"odel universes), i.e.,
\be
\epsilon_g = - \frac{\Omega^2}{8\pi} < 0
\ee
for the acausal model, and
\be
\epsilon_g = - \frac{\Omega^2}{4\pi} < 0
\ee
for the causal model. This can have an interesting connection to
the results of the calculation for brane universes \cite{barrow} where
it was shown that CTCs avoidance (and so the causality) is
possible provided the total effective energy density is {\it
negative} for these models.
Also, in both models all the components of the gravitational
Poynting 4-vector
\be
_g P^i = \left( \delta_k^i + v^i v_k \right) _E t_l^{~k} v^l
\ee
identically vanish in the Lorentzian coreper (\ref{eeee}). This
means that we have no gravitational energy flux which seems to be
related to the fact that the magnetic part of the Weyl (conformal curvature) tensor
vanishes for G\"odel models.

As far as the material part $T_{ik}$ of the canonical energy-momentum
complex (\ref{Ecomplex}) is concerned, its non-vanishing components
for the {\it acausal} G\"odel \cite{godel} model are \cite{paper1}
\be
\label{Tabacausal}
T_{00} =  \varrho + {\Lambda\over 8\pi} = \frac{\Omega^2}{8\pi},\hspace{0.5cm}
T_{11} = T_{22} = T_{33}
= - {\Lambda\over 8\pi} = \frac{\Omega^2}{8\pi},
\ee
and the following relation must be fulfilled ($\varrho$ - the
energy density of dust matter)
\begin{equation}
\label{rho}
4\pi \varrho = \Omega^2 = - \Lambda = {\rm const.}
\end{equation}
From (\ref{Tabacausal}) one can easily calculate that the matter
energy density
\be
\epsilon_m := T_{ik} v^i v^k = \frac{\Omega^2}{8\pi} > 0 ,
\ee
and that all the components of the material Poynting vector
\be
_m P^i := \left( \delta_k^{~i} + v^i v_k \right) T_l^{~k} v^l
\ee
identically vanish.
Combining the results for gravity and for matter we have
\be
\epsilon = \epsilon_g + \epsilon_m = 0 ,
\ee
and
\be
P^i := _g P^i + _m P^i = (0, 0, 0, 0)  ,
\ee
i.e., the total energy density and the total flux for the acausal model
{\it vanish}.
For the {\it causal} model we have \cite{paper1}
\be
\label{Tabcausal}
T_{00} =  \frac{e^2}{2} + {\Lambda\over 8\pi} = - \frac{\Omega^2}{8\pi},\hspace{0.5cm}
T_{11} = T_{22} = - \frac{e^2}{2} - {\Lambda\over 8\pi} = \frac{\Omega^2}{8\pi}, \hspace{0.5cm} T_{33}
= \frac{e^2}{2} - {\Lambda\over 8\pi} = \frac{3\Omega^2}{8\pi},
\ee
and the following relation between parameters $\Lambda, \Omega$ and $e$ has to be fulfilled
\be
\label{e}
\Lambda = -2 \Omega^2 = - 8\pi e^2 = {\rm const}.
\ee
From (\ref{Tabcausal}) there follows that
\be
\epsilon_m = - \frac{\Omega^2}{8\pi} , \hspace{0.3cm} _m P^i = (0,
0, 0, 0) .
\ee
This gives the result that the total energy density is {\it
negative} for the causal model and that its total flux vanishes, i.e.,
\bea
\epsilon & = & \epsilon_g  + \epsilon_m = - \frac{3\Omega^2}{8\pi} < 0 \\
P_i & = & _g P^i + _m P^i = (0, 0, 0, 0) .
\eea
All the above results look reasonable, but they valid only for the
Lorentzian frame (\ref{eeee}) and for a set of frame obtained
from it by the global Lorentz transformations. Then, one cannot
extract from them any coordinate-independent conclusions except for
matter part which can be transformed into an arbitrary frame by
tensorial transformation rule.

Finally, one can analyze the angular momentum of the G\"odel
spacetimes in the Lorentzian frame (\ref{eeee}) by using the
formulas (\ref{BTcomplex}), (\ref{FTangmom}), (\ref{gammas}) and (\ref{Riemmans}).
The calculations are simple, but somewhat tedious and this is why
we decided to put them into the Appendix. Here we only give some
general remarks.

At first, we would like to note that as many as 11 independent components in the acausal case and
13 independent components in the causal case of the Bergmann-Thomson angular momentum complex are
different from zero, and that it is difficult to extract any
more sophisticated physical conclusion from their shape. The only obvious conclusion is
that their non-vanishing reflects the fact of rotation of G\"odel
spacetimes. These remarks refer both to the gravitational part and
to the matter part of the Bergmann-Thomson complex.

Secondly, even after a decomposition of the Bergmann-Thomson angular
momentum complex into its tensor (t), vector
(v) and axial (a) (totally antisymmetric) parts as follows
\begin{equation}
\label{decomp}
M^{abc} = ^{(t)} M^{abc} + ^{(v)} M^{abc} + ^{(a)} M^{abc},
\end{equation}
where
\bea
^{(v)} M^{abc} := \frac{1}{3}\bigl(g^{bc}V^a - g^{ac} V^b\bigr),\\
^{(a)} M^{abc} = M^{[abc]} := \epsilon^{dabc}a_d,\\
V^a := M^{ab}_{~~b}, ~~a^d := - \frac{1}{6}
\epsilon^{dabc}M_{abc},\\
^{(t)} M^{abc} := M^{abc} - (^{(v)} M^{abc} + ^{(a)} M^{abc}) ,
\eea
the situation is still unclear, although much simpler. The reason is that we still have
too many non-vanishing independent components of these irreducible parts (12 for the
vectorial parts and 18 for the tensorial parts).

The same is true for the irreducible components of matter angular
momentum (\ref{FTangmom}) and gravitational angular momentum
\be
\label{grav}
_g M^{ika} = M^{ika} - _m M^{ika}
\ee
(except for axial parts of the matter angular momentum
densities which vanish).

\section{Conclusion}

In this paper we have analyzed energetic quantities for G\"odel
universes. In order to calculate these quantities we have used the
canonical energy-momentum pseudotensor/complex of Einstein and canonical
angular momentum pseudotensor/complex of Bergmann and Thomson. We have presented
these objects in an anholonomic Lorentzian frame and performed
the calculations in such an anholonomic frame which
substantially simplified the calculations.

We have found that for both considered acausal and causal G\"odel models, the Einstein
pseudotensor is traceless, not symmetric, and that the gravitational energy ``density'' is
negative. Also, the gravitational Poynting vector vanishes for
these models which seems to have a direct relation to the fact that the magnetic part
of the Weyl (conformal curvature) tensor vanishes for G\"odel models.
On the other hand, the total (gravitational and matter) energy ``density''
for the acausal model is zero, while for the causal model it
is negative. This last statement is {\it in agreement} with the results
obtained for the superenergy density \cite{paper1} which we found supportive for our
earlier superenergetic investigations. Also, there exists a
puzzling conicidence with the result obtained recently for brane
universes \cite{barrow}, where the total effective energy density
for these models must be negative in order to get causality.

On the other hand, the canonical angular momentum Bergmann-Thomson complex
has so complicated structure that practically it is difficult to
extract any more sophisticated physical conclusion, except that it does not vanish
which reflects the fact of global rotation of G\"odel spacetimes.

Naturally, these conclusions are valid only in the Lorentzian frame
applied and in a globally Lorentz rotated frame obtained from
this.

The main problem is that the calculated complexes are not tensors
and due to this one is not able to extract any convincing physical
information in a coordinate-independent way. In particular, the
application of the Landau-Lifshitz and M\o ller pseudotensors in a holonomic frame for
the acausal G\"odel spacetime recently \cite{Gad}, shows that the
results obtained differ from ours.

In this context we emphasize that superenergetic quantities are tensors and so they admit a
coordinate-independent description of the gravitational field so
that the agreement of the results obtained for energetic
quantities with the results obtained for superenergetic quantities suggests also usefulness
of the Einstein and Bergmann-Thomson complexes in the analysis.
However, from what we derived, it appears that the analytic structure of the
canonical superenergy tensors
and the canonical supermomentum tensors for matter and gravitation
is {\it much simpler} than the analytic structure of the corresponding
canonical energetic pseudotensors/complexes.

%\section{Acknowledgments}

\appendix

\section{Bergmann-Thomson angular momentum complex components for G\"odel
universes}

\setcounter{equation}{0}

From (\ref{Tabacausal}) and (\ref{Tabcausal}) we can calculate the non-vanishing components of the matter angular
momentum tensor (\ref{FTangmom}) which is antisymmetric in the first two indices.
For the acausal model they can be presented in a compact way as follows
\bea
\label{Macausal}
_m M^{0\mu\mu} & = & - _m M^{\mu 0 \mu} = \frac{r^0 \Omega^2}{8 \pi}
, \nonumber \\
_m M^{0\mu 0} & = & - _m M^{\mu 0 0} = - \frac{r^{\mu} \Omega^2}{8 \pi}
, \nonumber \\
_m M^{\mu\nu\mu} & = & - _m M^{\nu\mu\mu} = - \frac{r^{\nu} \Omega^2}{8 \pi}
,
\eea
and the Greek indices $\mu, \nu, \ldots = 1, 2, 3$, $\mu \neq \nu$. For the
causal model one can use somewhat less compact way of presentation, i.e.,
\bea
\label{Mcausal}
_m M^{0 \mu 0} & = & - _m M^{\mu 0 0} = \frac{r^{\mu} \Omega^2}{8 \pi}
, \nonumber \\
_m M^{011} & = & - _m M^{101} = _m M^{022} = - _m M^{202} =
\frac{r^0\Omega^2}{8\pi} , \nonumber \\
_m M^{033} & = & - _m M^{303} = \frac{3 r^0\Omega^2}{8\pi},
\nonumber \\
_m M^{122} & = & - _m M^{212} = \frac{r^1\Omega^2}{8\pi} , \nonumber \\
_m M^{133} & = & - _m M^{313} = \frac{3r^1\Omega^2}{8\pi} , \nonumber \\
_m M^{233} & = & - _m M^{323} = \frac{3r^2\Omega^2}{8\pi} , \nonumber \\
_m M^{121} & = & - _m M^{211} = - \frac{r^2\Omega^2}{8\pi} , \nonumber \\
_m M^{131} & = & - _m M^{311} = _m M^{232}  =  - _m M^{322} = - \frac{r^3\Omega^2}{8\pi}
.
\eea

As for the Bergmann-Thomson complex for the sake of performing the exact calculations
we split the formula (\ref{BTcomplex}) as follows
\be
_{BT} M^{ika} = A^{ika} + B^{ika} + C^{ika} + D^{ika} + E^{ika} ,
\ee
where
\bea
\label{ABCDE}
A^{ika} & = & \frac{1}{2\chi} \left[ \eta_{lmrn} g^{tr}
\gamma^{m}_{tp} \left( \eta^{aipn} g^{kl} - \eta^{akpn} g^{il}
\right) \right] ,\\
B^{ika} & = & \frac{1}{2\chi} \left[
\eta^{atbs} \eta_{lmrs} \gamma^{m}_{nb} g^{nr} r^p
\left( g^{il} \gamma^{k}_{pt} - g^{kl} \gamma^{i}_{pt} \right)
\right] ,\\
C^{ika} & = & \frac{1}{2\chi}
\left( r^i g^{kl} - r^k g^{il} \right) \eta^{aspn}
\eta_{tjmn} g^{rm} \gamma^{t}_{ls} \gamma^{j}_{rp}  ,\\
D^{ika} & = & - \frac{1}{2\chi}
\left( r^i g^{kl} - r^k g^{il} \right) \eta^{aspn}
\eta_{ljmn} g^{rm} \gamma^{t}_{rs} \gamma^{j}_{tp} ,\\
E^{ika} & = & - \frac{1}{4\chi}
\left( r^i g^{kl} - r^k g^{il} \right) \eta^{aspn}
\eta_{ljms}
g^{tm} R^{j}_{~tnp} .
\eea
Taking into account the Equations (\ref{gammas}) and
(\ref{Riemmans}) one gets for (\ref{ABCDE}) the following expressions
\bea
2 \chi A^{ika} = & - & m \sqrt{2} \left(\eta^{ai23}g^{k2} -
\eta^{ak23}g^{i2}\right) - m \sqrt{2} \left(\eta^{ai13}g^{k1} -
\eta^{ak13}g^{i1} \right) \nonumber \\
& - & m \sqrt{2} \left(\eta^{ai03}g^{k0} -
\eta^{ak03}g^{i0} \right) + 2m \left(\eta^{ai23}g^{k0} -
\eta^{ak23}g^{i0} \right) \nonumber \\
& - & 2m \left(\eta^{ai20}g^{k3} -
\eta^{ak20}g^{i3} \right) ,
\eea
\bea
2 \chi B^{ika} = & - & m\sqrt{2} \eta^{a023} r^1 \left( g^{i2}
\gamma^{k}_{10} - g^{k2} \gamma^{i}_{10} \right) -
m\sqrt{2} \eta^{a023} r^2 \left( g^{i2}
\gamma^{k}_{20} - g^{k2} \gamma^{i}_{20} \right) \nonumber \\
& - & m\sqrt{2} \eta^{a123} r^0 \left( g^{i2}
\gamma^{k}_{01} - g^{k2} \gamma^{i}_{01} \right) - m\sqrt{2} \eta^{a123} r^2 \left( g^{i2}
\gamma^{k}_{21} - g^{k2} \gamma^{i}_{21} \right) \nonumber \\
& - & m\sqrt{2} \eta^{a013} r^1 \left( g^{i1}
\gamma^{k}_{10} - g^{k1} \gamma^{i}_{10} \right) -
m\sqrt{2} \eta^{a013} r^2 \left( g^{i1}
\gamma^{k}_{20} - g^{k1} \gamma^{i}_{20} \right) \nonumber \\
& - & m\sqrt{2} \eta^{a213} r^2 \left( g^{i1}
\gamma^{k}_{22} - g^{k1} \gamma^{i}_{22} \right) -
m\sqrt{2} \eta^{a213} r^1 \left( g^{i1}
\gamma^{k}_{12} - g^{k1} \gamma^{i}_{12} \right) \nonumber \\
& - & m\sqrt{2} \eta^{a213} r^0 \left( g^{i1}
\gamma^{k}_{02} - g^{k1} \gamma^{i}_{02} \right) -
m\sqrt{2} \eta^{a103} r^0 \left( g^{i0}
\gamma^{k}_{01} - g^{k0} \gamma^{i}_{01} \right) \nonumber \\
& - & m\sqrt{2} \eta^{a103} r^2 \left( g^{i0}
\gamma^{k}_{21} - g^{k0} \gamma^{i}_{21} \right) -
m\sqrt{2} \eta^{a203} r^2 \left( g^{i0}
\gamma^{k}_{22} - g^{k0} \gamma^{i}_{22} \right) \nonumber \\
& - & m\sqrt{2} \eta^{a203} r^1 \left( g^{i0}
\gamma^{k}_{12} - g^{k0} \gamma^{i}_{12} \right) -
m\sqrt{2} \eta^{a203} r^0 \left( g^{i0}
\gamma^{k}_{02} - g^{k0} \gamma^{i}_{02} \right) \nonumber \\
& + & 2m \eta^{a023} r^1 \left( g^{i0}
\gamma^{k}_{10} - g^{k0} \gamma^{i}_{10} \right) +
2m \eta^{a023} r^2 \left( g^{i0}
\gamma^{k}_{20} - g^{k0} \gamma^{i}_{20} \right) \nonumber \\
& + & 2m \eta^{a123} r^0 \left( g^{i0}
\gamma^{k}_{01} - g^{k0} \gamma^{i}_{01} \right) +
2m \eta^{a123} r^2 \left( g^{i0}
\gamma^{k}_{21} - g^{k0} \gamma^{i}_{21} \right) \nonumber \\
& - & 2m \eta^{a120} r^0 \left( g^{i3}
\gamma^{k}_{01} - g^{k3} \gamma^{i}_{01} \right) -
2m \eta^{a120} r^2 \left( g^{i3}
\gamma^{k}_{21} - g^{k3} \gamma^{i}_{21} \right) ,
\eea
\bea
2 \chi C^{ika} = & - & 2m^2 \left( r^i g^{k2} - r^k g^{i2} \right)
\eta^{a013} - 2m^2 \left( r^i g^{k0} - r^k g^{i0} \right)
\eta^{a213} \nonumber \\
& + & 2m^2 \left( r^i g^{k1} - r^k g^{i1} \right) \eta^{a023} -
2m^2 \sqrt{2} \left( r^i g^{k2} - r^k g^{i2} \right) \eta^{a123} ,
\eea
\bea
2 \chi D^{ika} & = & m^2 \left( r^i g^{k2} - r^k g^{i2} \right)
\eta^{a123} + m^2 \left( r^i g^{k2} - r^k g^{i2} \right)
\eta^{a013} \nonumber \\
& + & m^2 \left( r^i g^{k1} - r^k g^{i1} \right)
\eta^{a203} + m^2 \left( r^i g^{k0} - r^k g^{i0} \right)
\eta^{a213} \nonumber \\
& + & m^2 \left( r^i g^{k3} - r^k g^{i3} \right)
\eta^{a021} ,
\eea
\bea
2 \chi E^{ika} & = & 2 \Omega^2 \left( r^i g^{k2} - r^k g^{i2} \right)
\eta^{a301} - 2 \Omega^2 \left( r^i g^{k1} - r^k g^{i1} \right)
\eta^{a302} \nonumber \\
& - & 2 \left( 3\Omega^2 - m^2 \right) \left( r^i g^{k0} - r^k g^{i0} \right)
\eta^{a312} + 2 \left( \Omega^2 - m^2 \right)  \left( r^i g^{k3} - r^k g^{i3} \right)
\eta^{a012}
\eea

Finally, we present the non-vanishing components of the Bergmann-Thomson complex
(\ref{BTcomplex}) which contain both matter and gravitation (remember they are antisymmetric
in the first two indices). These are:
\bea
\label{TotalBT}
_{BT} M^{010} & = & \frac{(3\Omega^2 - m^2)r^1 + m}{8\pi},
\nonumber \\
_{BT} M^{020} & = & \frac{(3\Omega^2 - m^2)r^2}{8\pi},
\nonumber \\
_{BT} M^{030} & = & \frac{(6\Omega^2 - m^2) r^3}{16\pi},
\nonumber \\
_{BT} M^{230} & = & \frac{m^2 \sqrt{2} r^3}{16\pi},
\nonumber \\
_{BT} M^{011} & = & \frac{\Omega^2 r^0}{8\pi},
\nonumber \\
_{BT} M^{121} & = & - \frac{\Omega^2 r^2}{8\pi},
\nonumber \\
_{BT} M^{131} & = & \frac{(m^2 - 2 \Omega^2)r^3}{16\pi},
\nonumber \\
_{BT} M^{022} & = & \frac{\Omega^2 r^0}{8\pi},
\nonumber \\
_{BT} M^{122} & = & \frac{\Omega^2 r^1}{8\pi},
\nonumber \\
_{BT} M^{232} & = & \frac{(m^2 - 2 \Omega^2)r^3}{16\pi},
\nonumber \\
_{BT} M^{033} & = & \frac{(3m^2 - 2\Omega^2)r^0 + m^2 \sqrt{2}
r^2}{16\pi}, \nonumber \\
_{BT} M^{133} & = & \frac{2m + (3m^2 - 2\Omega^2)r^1}{16\pi},
\nonumber \\
_{BT} M^{233} & = & \frac{(3m^2 - 2\Omega^2)r^2 + m^2 \sqrt{2}
r^0}{16\pi} .
\eea
As one can see, for the acausal model $(m^2 = 2\Omega^2)$ one has
11 independent components and for the causal model $(m^2 = 4\Omega^2)$
there are 13 independent components.

Subtracting matter angular momentum from the Bergmann-Thomson
complex we can obtain the components of the gravitational angular
momentum pseudotensor (\ref{grav}) for both models. Applying (\ref{Macausal}) and
(\ref{TotalBT}) for the acausal model we have 9 non-vanishing components
\bea
_{g} M^{010} & = & \frac{\sqrt{2} \Omega}{8 \pi} + \frac{\Omega^2}{4\pi} r^1   ,
\nonumber \\
_{g} M^{020} & = & \frac{\Omega^2}{4\pi} r^2  ,
\nonumber \\
_{g} M^{030} & = & \frac{3\Omega^2}{8\pi} r^3 ,
\nonumber \\
_{g} M^{230} & = & \frac{\sqrt{2} \Omega^2}{8\pi} r^3 , \nonumber \\
_{g} M^{131} & = & \frac{\Omega^2}{8\pi} r^3,
\nonumber \\
_{g} M^{232} & = & \frac{\Omega^2}{8\pi} r^3,
\nonumber \\
_{g} M^{033} & = & \frac{\Omega^2}{8\pi} r^0 + \frac{\sqrt{2}\Omega^2}{8\pi} r^2,
\nonumber \\
_{g} M^{133} & = & \frac{\sqrt{2} \Omega}{8\pi} + \frac{\Omega^2}{8\pi} r^1 ,
\nonumber \\
_{g} M^{233} & = & \frac{\Omega^2}{8\pi} r^2 + \frac{\sqrt{2}\Omega^2}{8\pi} r^0 .
\eea
while using (\ref{Mcausal}) and (\ref{TotalBT}) for the causal model we have
8 independent components
\bea
_{g} M^{010} & = &  \frac{\Omega}{4 \pi} - \frac{\Omega^2}{4\pi} r^1  ,
\nonumber \\
_{g} M^{020} & = & - \frac{\Omega^2}{4\pi} r^2  ,
\nonumber \\
_{g} M^{230} & = & \frac{\sqrt{2} \Omega^2}{4\pi} r^3 , \nonumber \\
_{g} M^{131} & = & - \frac{\Omega^2}{4\pi} r^3,
\nonumber \\
_{g} M^{232} & = & - \frac{\Omega^2}{4\pi} r^3,
\nonumber \\
_{g} M^{033} & = & \frac{\Omega^2}{4\pi} r^0 + \frac{\sqrt{2}\Omega^2}{4\pi} r^2,
\nonumber \\
_{g} M^{133} & = &   \frac{\Omega}{4\pi} + \frac{\Omega^2}{4\pi} r^1 ,
\nonumber \\
_{g} M^{233} & = & \frac{\Omega^2}{4\pi} r^2 + \frac{\sqrt{2}\Omega^2}{4\pi} r^0 .
\eea

\end{document}